\documentclass[10pt]{ismd08}
\usepackage{graphicx}

\usepackage{wrapfig}
\usepackage{amssymb}
\usepackage{amsmath}
\usepackage{units}
\usepackage{cite,./mcite}

\newcommand{\pb}{\ensuremath{\:\textrm{pb}^{-1}}}
\newcommand{\fb}{\ensuremath{\:\textrm{fb}^{-1}}}
\newcommand{\nb}{\ensuremath{\:\textrm{nb}}}
\newcommand{\GeV}{\ensuremath{\:\textrm{Ge\kern -0.1em V}}}
\newcommand{\MeV}{\ensuremath{\:\textrm{Me\kern -0.1em V}}}
\newcommand{\KeV}{\ensuremath{\:\textrm{ke\kern -0.1em V}}}
\newcommand{\eV}{\ensuremath{\:\textrm{e\kern -0.1em V}}}
\newcommand{\PYTHIA}{\textsc{Pythia}}
\newcommand{\pslash}{p\llap{/\kern+0.1em}}

\begin{document}
\title{Heavy Quark Production at HERA as a Probe of Hard QCD}
\author{R. Shehzadi (on behalf of the H1 and ZEUS Collaborations)}
\institute{{\it Physikalisches Institut, Universit\"at Bonn, Germany}}
\maketitle
\begin{abstract}
The study of heavy flavor production is a central topic of research at HERA and is an important testing ground for perturbative QCD. A selection of results for charm and beauty production in $\gamma p$, using different experimental techniques and compared to different theoretical predictions, obtained by the H1 and ZEUS collaborations will be presented. 
\end{abstract}

\section{Introduction}

Heavy flavor production at HERA is an important tool to investigate our present understanding of the theory of Quantum Chromodynamics (QCD). In $e^{\pm}p$ collisions, the main production mechanism for heavy flavors is the Boson Gluon Fusion (BGF) process. The large mass of the heavy quarks produced in this process provides a hard scale so that calculations in perturbative QCD are expected to be reliable. However, the simultaneous presence of competing hard scales, such as the transverse momentum ($p_T$) of the heavy quark or the virtuality of the exchanged photon ($Q^2$) induces additional theoretical uncertainties due to terms in the perturbative expansion which depend logarithmically on the ratio of these scales. Since the perturbative expansion can not be optimized for all scales at once, different calculational approaches have been developed assuming a single hard scale in each. Therefore, comparisons of the measured cross sections with theory predictions are particularly sensitive to the way the perturbative expansion is made.

Different kinematic variables are used to describe the $ep$ interaction at HERA:  the photon's virtuality $Q^2$, the Bjorken scaling variable, $x$, and the inelasticity, $y$. Until 1997 HERA ran at a centre-of-mass energy of $\sqrt{s}=300\GeV$. The proton energy was increased leading to 
$\sqrt{s}=320\GeV$ for data taken from 1998 onwards. During a shutdown in 2000 and 2001 the accelerator and detectors were upgraded. The period up to 2000 is usually called HERA-I and after 2000 HERA-II. By the end of the running both of the colliding-beam experiments, H1 and ZEUS, had collected about 0.5$\fb$ of data.

The kinematic range of the analyzed data can be separated into the following two regimes: photoproduction ($\gamma p$), where the exchanged photon in the
process is almost real, and deep inelastic scattering (DIS), where the
exchanged photon is virtual. Experimentally,
$\gamma p$ is defined by the scattered electron not being in
the acceptance region of the main detectors, corresponding to a cut $Q^2 \lesssim 1 \GeV^2 $.
In the following a small selection of recent measurements of heavy quark production in $\gamma p$ will be presented. 

\section{Theoretical Models}
\label{sec:theory}

For heavy flavor photoproduction at HERA, different possible theoretical schemes have been used. These include:\\
\begin{itemize}
\vspace {-15pt} 
\item The leading order (LO) plus parton shower approach, where leading order QCD matrix elements are complemented by parton showers. This approach is implemented in many Monte Carlo (MC) models, e.g. \PYTHIA~\cite{pythia}, which is based on collinear factorization and DGLAP~\cite{dglap} evolution of parton densities, and CASCADE~\cite{cascade} based on $k_T$ factorization~\cite{kt} which uses a $k_T$ unintegrated gluon density that is evolved according to CCFM~\cite{ccfm} evolution. These MC models are mostly used for acceptance corrections.
\item The next-to-leading order (NLO) massive approach~\cite{fmnr}. This approach assumes that there is no intrinsic charm or beauty in the proton (or photon). The heavy quarks are only produced dynamically in the hard scattering. This approach is expected to work best when all relevant hard scales e.g. the quark's transverse momentum $p_T$  are of the order of heavy quark mass ($m_q$). This scheme is also known as fixed order (FO) scheme.
\item The NLO massless approach. For $p_T \gg m_q$, large $\log$($p_T/m_q$) terms could in principle spoil the reliability of the predictions. In this case, it might be preferable to switch to massless scheme, in which the $m_q$ is neglected kinematically. The potentially large logarithms can then be resummed to all orders (next-to-leading log (NLL) resummation). Since such an approach is obviously not applicable when $p_T \sim m_q$, schemes have been designed which make a continuous transition between the FO massive and NLL massless scheme. This is often referred to as the GM-VFN (Generalized Mass Variable Flavor Number) scheme~\cite{massive}.
\end{itemize} 
\section{Experimental methods}

\label{sec:exp}
On the experimental side, there are several different
methods to tag the heavy quark final state. Different methods often cover different kinematic ranges.
The charm-quark events are frequently tagged by the presence of $D^{*}$ mesons. The measurement of beauty-quark events is difficult due to the fact that beauty production is suppressed with respect to charm production by the large $b$ mass and by its smaller coupling to the photon.
Two basic techniques are used to tag beauty events: The measurement of track impact parameters ($\delta$), which enrich beauty production due to the large lifetime of beauty hadrons, and measurements based on semileptonic decays of the $b$ quark. In the latter case, the large momentum of the lepton transverse to the direction of $b$-initiated jet ($p_{T}^{rel}$), due to the large $b$ mass, is used to discriminate against semileptonic charm decays or misidentified light flavor events.
Finally, a lepton tag can e.g. be combined with a lifetime tag or with a second lepton tag. The double lepton tag enables one to go to lower transverse momenta.\\
The above mentioned methods have been used by both the H1 and ZEUS collaborations.
\section{Charm production}

\begin{wrapfigure}[7]{r}{0.39\textwidth}
  \vspace{-38pt}
  \begin{center}
    \includegraphics[width=0.37\textwidth, clip=true]{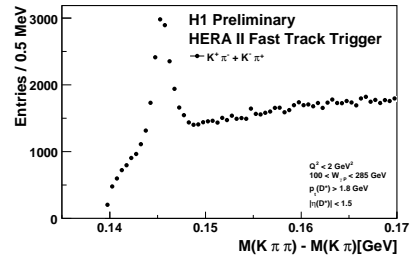}
\end{center}
 \vspace{-28pt}
\caption{The distribution of $\Delta M$ in $\gamma p$ events}
\label{deltaM}
\end{wrapfigure}
\begin{figure}[htbp]
\vspace {-8pt}
   \begin{center}
     \begin{tabular}{cc}
       \begin{minipage}{0.5\textwidth}
         \begin{center}
           \includegraphics[width=1.\textwidth, clip=true]{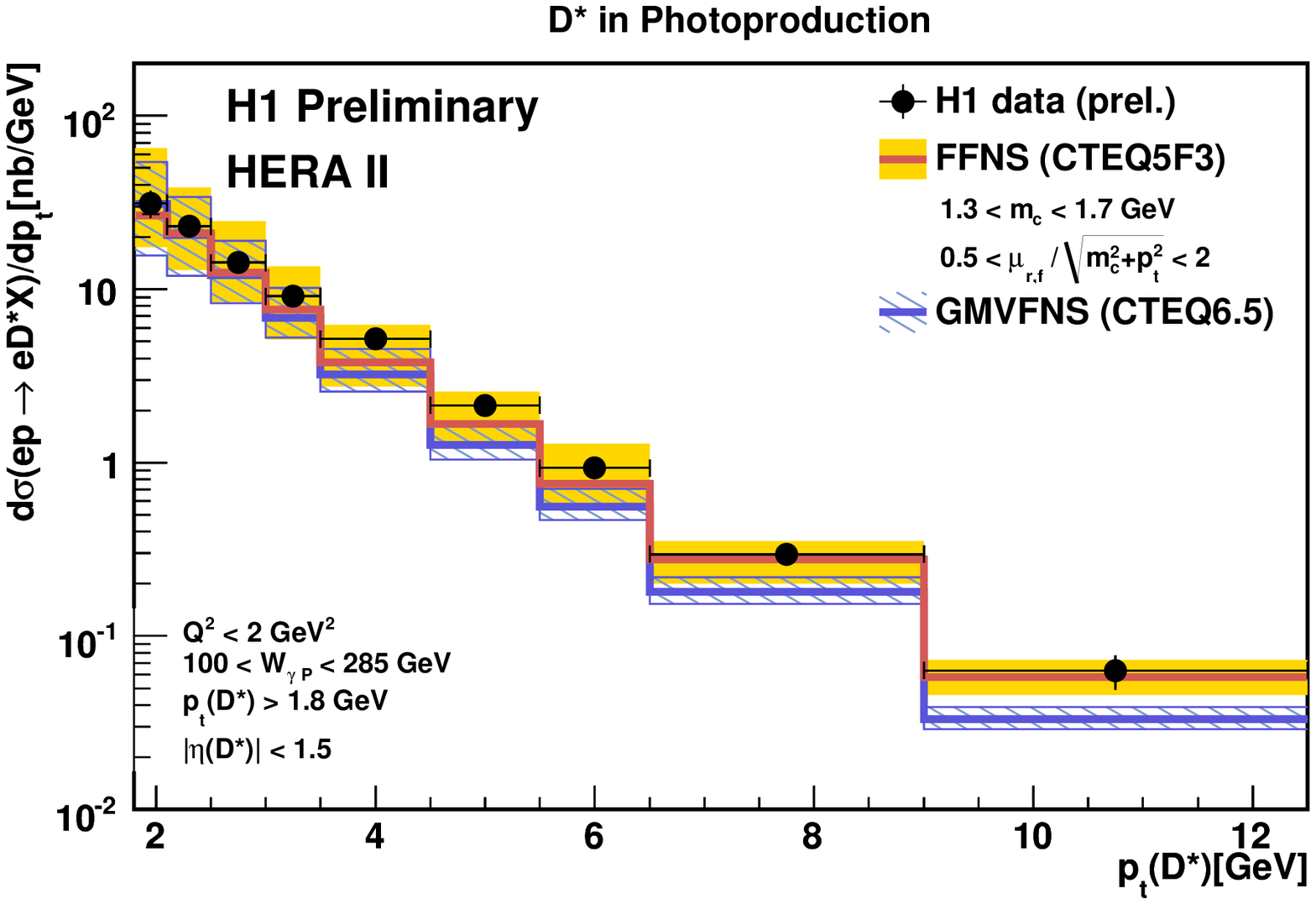}
         \end{center}
       \end{minipage}
       \begin{minipage}{0.5\textwidth}
         \begin{center}
           \includegraphics[width=1.\textwidth, clip=true]{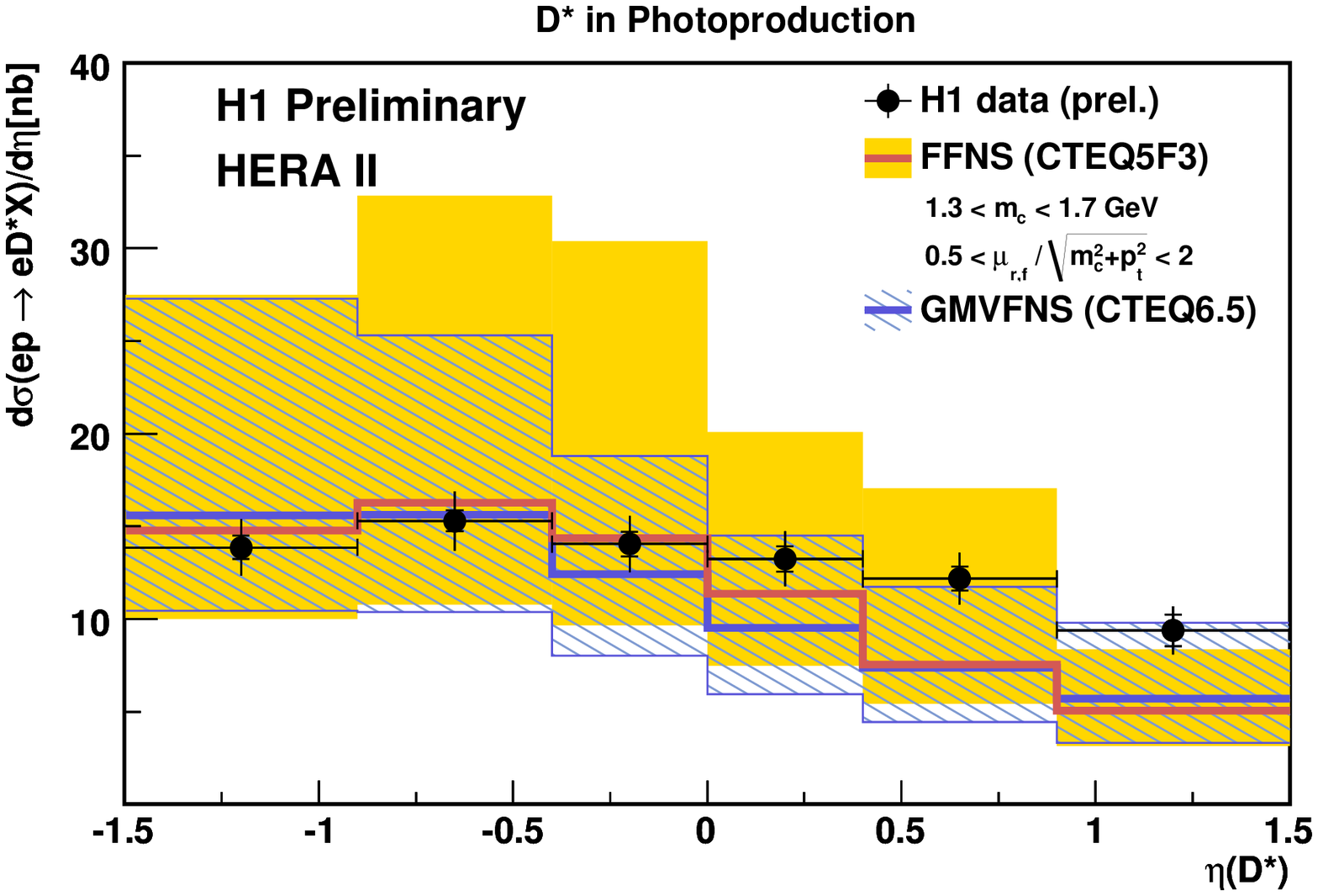}
         \end{center}
       \end{minipage}
     \end{tabular}
   \end{center}
  \vspace{-21pt}
   \caption{The H1 measurement of $D^{*}$ mesons in $\gamma p$ interactions, compared to two QCD models at NLO: the FMNR program in the FFNS (shaded area) and a new calculation in the GM-VFNS (hatched)}
   \label{pt_eta}
\end{figure}

The H1 collaboration has recently released a new measurement of charm photoproduction~\cite{H1D*}.
This measurement is based on HERA-II data corresponding to an integrated luminosity of 93$\pb$ .
The charm events are tagged by the presence of a $D^{*}$ meson decaying in the so-called golden channel $D^{* \pm} \rightarrow D^{0} \pi^{\pm}$, $D^{0} \rightarrow K^{-} \pi^{+}$. For this, they make use of their new fast track trigger which enables events with $D^{*}$ candidates to be selected early in the trigger chain and allows to reconstruct their mass. A clear signal is seen as illustrated in Figure~\ref{deltaM}. The kinematic cuts are also indicated in the figure. \\
Differential cross sections as a function of $p_T(D^{*})$, $\eta (D^{*})$ and $W(D^{*})$ as well as double differential cross sections in $p_T(D^{*})$ and $\eta(D^{*})$ have been measured and compared to various QCD models.
Differential distributions in $p_T(D^{*})$ and $\eta(D^{*})$ compared to two models at NLO QCD are shown in Figure~\ref{pt_eta}.
The two models are FMNR~\cite{fmnr}, which is based on the massive scheme and GM-VFNS~\cite{vfns}, which uses the combined scheme. Both models show similar behavior. The $p_T(D^{*})$ spectrum is well described with a slight deficiency at high $p_T$ in the case of GM-VFNS model. For $\eta(D^{*})$ both predictions have a somewhat different shape compared to that of the data and theoretical uncertainties are several times larger than the experimental ones.
\section{Beauty production}

The H1 and ZEUS collaborations have recently reported new measurements based on lepton tags.
\begin{figure}[htbp]
\vspace {-13pt}
   \begin{center}
     \begin{tabular}{cc}
       \begin{minipage}{0.5\textwidth}
         \begin{center}
           \includegraphics[width=1.\textwidth, clip=true]{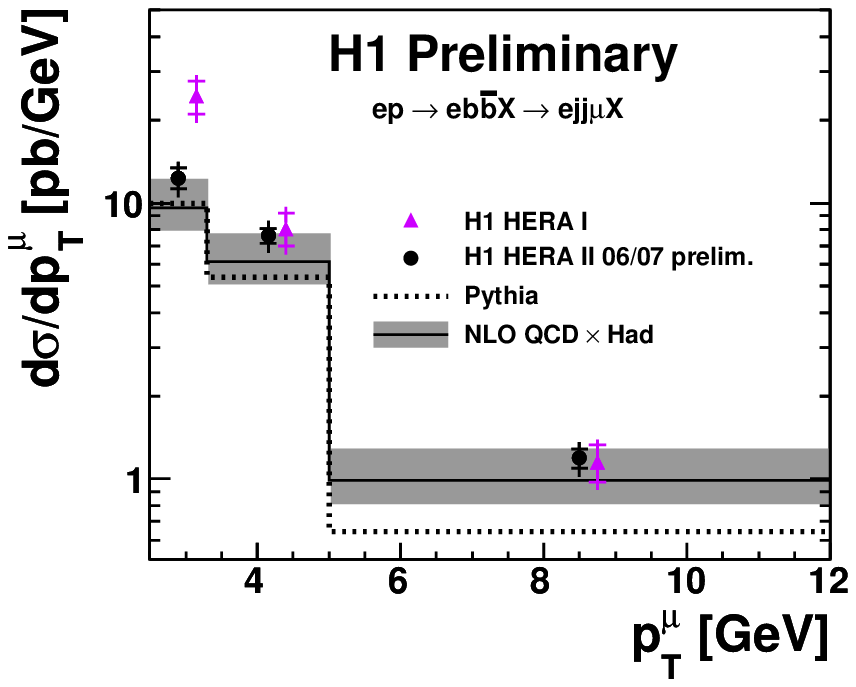}
         \end{center}
       \end{minipage}
       \begin{minipage}{0.5\textwidth}
         \begin{center}
           \includegraphics[width=.75\textwidth, clip=true]{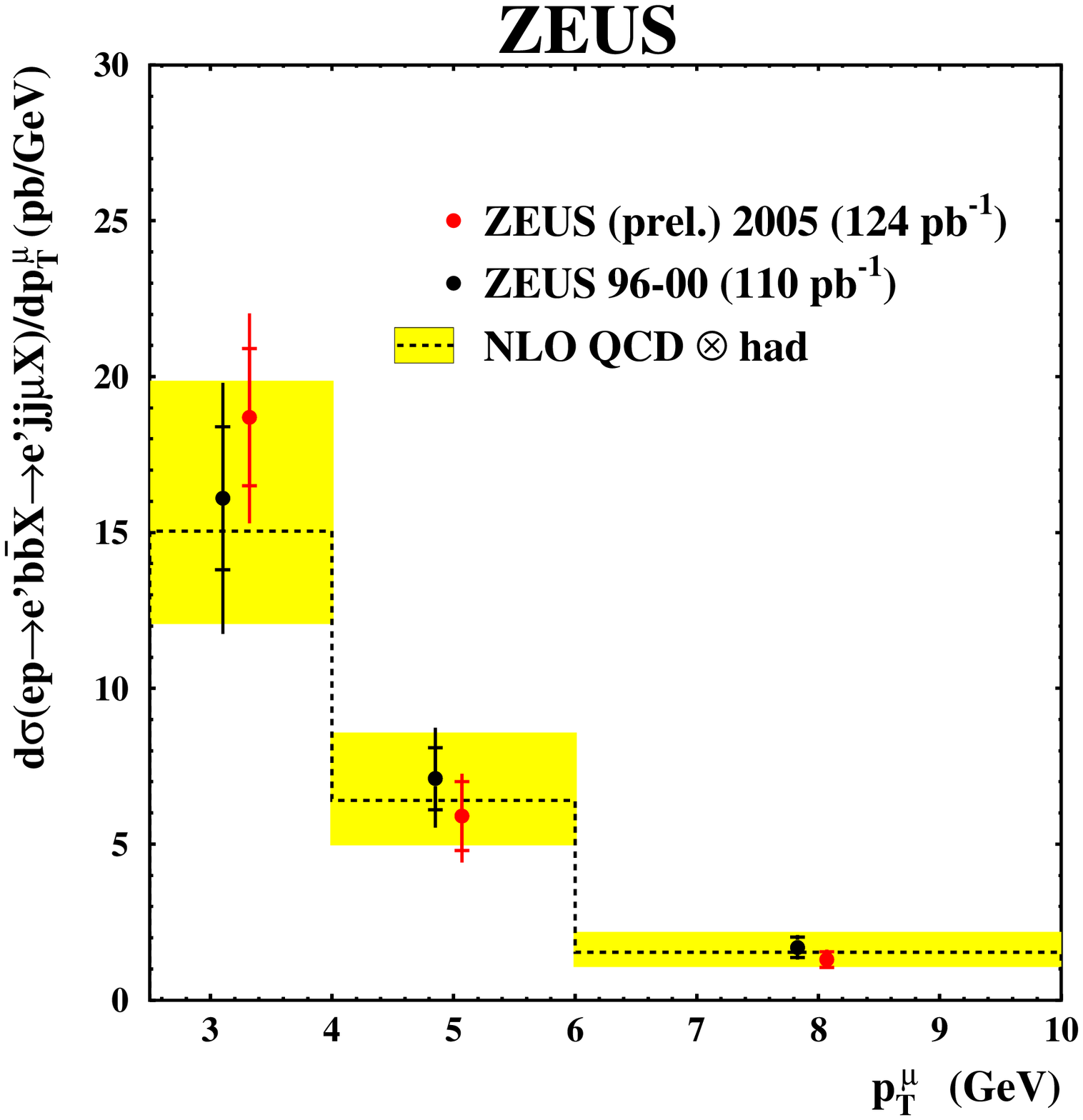}
         \end{center}
       \end{minipage}
     \end{tabular}
   \end{center}
\vspace {-21pt}
   \caption{b-quark production cross-section in $\gamma p$ as a function of the transverse momentum of the muon $p_T^{\mu}$: result from the H1 collaboration (left), result from the ZEUS collaboration (right).}
   \label{pt_mu}
\end{figure} 
The two measurements, one from H1~\cite{H1mu}, and one from ZEUS~\cite{zeusmu} use semileptonic decays to muons to identify heavy quark decays.
For this a sample of dijets events requiring at least two jets with $|\eta^{jet1(2)}| < 2.5$ and jet transverse momentum above $p_T^{jet1(2)} > 7(6)\GeV$ is used. The fraction of beauty events is then determined from fits to the distributions of $p_T^{rel}$ of the muon with respect to jet axis and the impact parameter $\delta$ of the muon track.
Both the H1 and ZEUS collaborations use HERA-II data corresponding to integrated luminosities 171$\pb$ and 124$\pb$, respectively.
Both measurements cover the range of $0.2 < y < 0.8$, $Q^2 < 1\GeV^{2}$, muon transverse momentum $p_T^{\mu} > 2.5\GeV$; however the ZEUS data cover a range $-1.6 < \eta^{\mu} < 2.3$ of muon pseudorapidity while the H1 analysis is restricted to $-0.55 < \eta^{\mu} < 1.1$. The cross section in $p_T^{\mu}$ for the two analyses is shown in Figure~\ref{pt_mu}.
Both analyses show good agreement with perturbative QCD calculations performed with the FMNR program. The excess of data over NLO predictions at low values of jet and muon transverse momentum that was reported in an earlier H1 analysis of HERA-I data~\cite{H1hera1} is not confirmed by the new H1 analysis.

The ZEUS collaboration has made another measurement based on the identification of both heavy quark decays~\cite{dimuon}. This measurement uses HERA-I data corresponding to an integrated
 luminosity of 114$\pb$. Events with two muons in the final state are selected.
Because of the
\begin{wrapfigure}[28]{r}{0.501\textwidth}
 \vspace{-14pt}
  \begin{center}
    \includegraphics[width=.45\textwidth, clip=true]{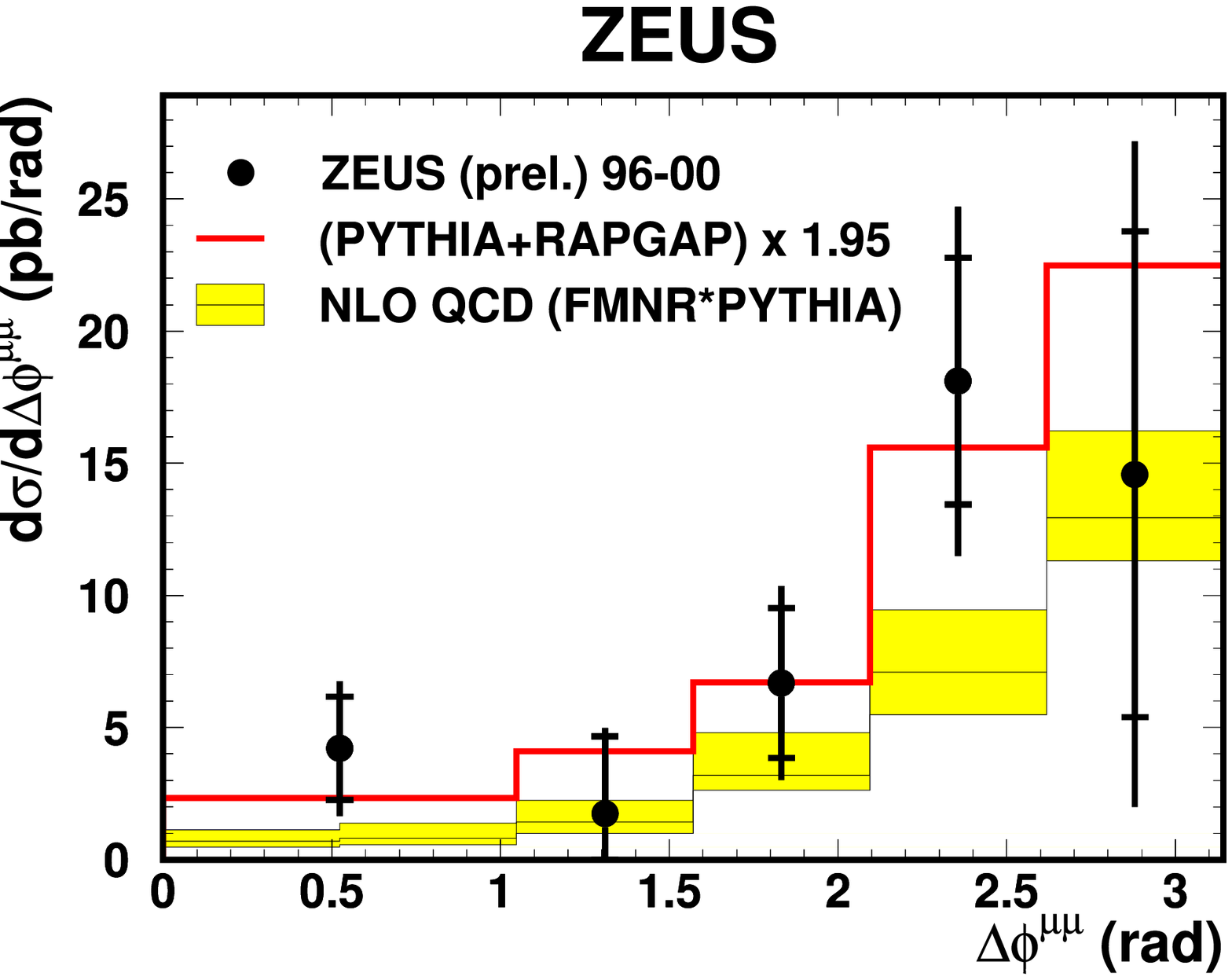}
  \end{center}
 \vspace{-20pt}
  \caption{The dimuon cross-section as a function of the azimuthal angle between the muon in dijet events.}
  \label{dimuon}
 \vspace{15pt}
 \includegraphics[width=0.497\textwidth, clip=true]{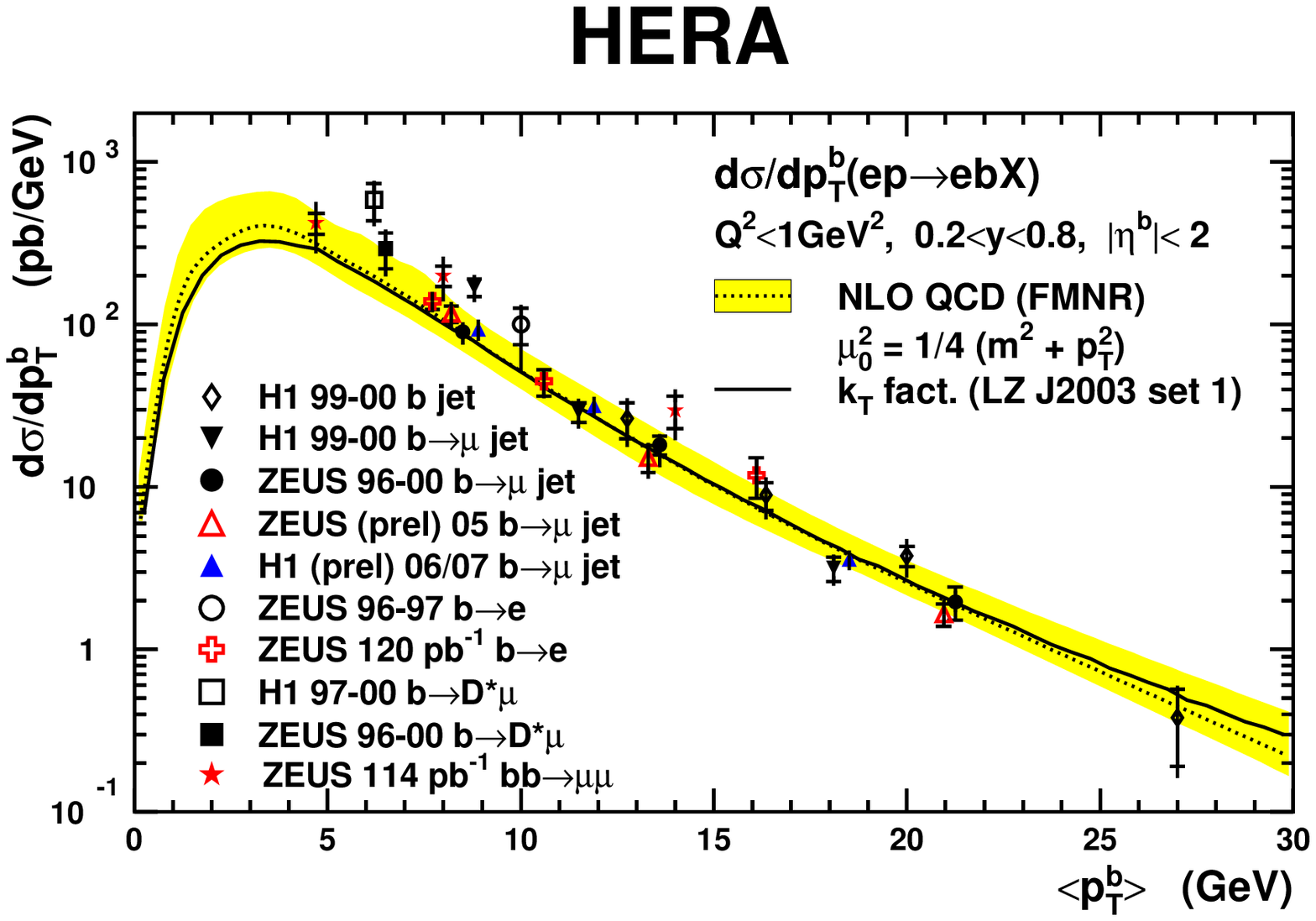}
 \vspace{-20pt}
\caption{Cross sections for beauty production in $\gamma p$ at HERA as a function of $p_T^{b}$}
\label{summary}
\end{wrapfigure}
high beauty fraction in such a dimuon sample, jets are not required in the selection.
This selection of double tags has several advantages: a larger kinematic range is accessible
and the background is reduced substantially. It allows the measurement of $b\bar{b}$ correlations, which probe the next-to-leading order effects.
A low $p_T$ threshold for muon identification, $p_T^{\mu} > 1.5 \GeV$ or even $p_T^{\mu} > 0.75 \GeV$ for high quality muon candidates and large rapidity coverage makes it possible to measure the total beauty production cross section with relatively small extrapolation.
The total cross section for the process $ep \rightarrow b\bar{b} X$ at $\sqrt{s} = 318 \GeV$ has been determined to be $\sigma = 13.9 \pm 1.5 (stat.)~^{+4.0}_{-4.3}(syst.) \nb$ to be compared to the NLO QCD prediction of $\sigma^{NLO}_{tot}(ep \rightarrow b\bar{b} X) = 7.5~^{+4.5}_{-2.1} \nb$. Within the large uncertainties, in particular of the NLO calculation,  the NLO prediction is consistent with the data.
Differential cross sections and measurements of $b\bar{b}$ correlations are also obtained and compared to other beauty cross section measurements, MC models and NLO QCD predictions. The distribution of the extracted cross section compared to NLO QCD prediction and leading order MC is shown in Figure~\ref{dimuon}. Both predictions agree with the data.\\
Figure~\ref{summary} summarizes all recent HERA measurements of $b$ production in $\gamma p$ as a function of $b$ quark transverse momentum ($p_T^{b}$). The different measurements agree well with each other and are in reasonable agreement with the NLO predictions.  
\section{Conclusion}

The study of heavy flavor production in $\gamma p$ at HERA remains a source of interest for testing and understanding the perturbative QCD. Several results based on HERA-II data set are now available. Some of these recent results using different tagging methods were presented. The charm results are in good agreement with NLO QCD. The beauty measurements are also in reasonable agreement with NLO QCD. The uncertainty on experimental results in both beauty and charm production are typically smaller than theoretical uncertainties.


\end{document}